# High accuracy capillary network representation in digital rock reveals permeability scaling functions

*Rodrigo F. Neumann[1\*], Mariane Barsi-Andreeta[2], Everton Lucas-Oliveira[2], Hugo Barbalho[1†], Willian A. Trevizan[3], Tito J. Bonagamba[2] and Mathias Steiner[1\*]*

[1] *IBM Research, Rio de Janeiro, RJ, 22290-240, Brazil*
[2] *São Carlos Institute of Physics, University of São Paulo, PO Box 369, São Carlos, SP, 13560-970, Brazil*
[3] *CENPES/Petrobras, Rio de Janeiro, RJ, 21941-915, Brazil*
[†] *Present address: Dell EMC R&D Center, Rio de Janeiro, RJ, 21941-907, Brazil*
[\*] *Corresponding authors:* rneumann@br.ibm.com *and* mathiast@br.ibm.com

## Abstract

Permeability is the key parameter for quantifying fluid flow in porous rocks. Knowledge of the spatial distribution of the connected pore space allows, in principle, to predict the permeability of a rock sample. However, limitations in feature resolution and approximations at microscopic scales have so far precluded systematic upscaling of permeability predictions. Here, we report fluid flow simulations in capillary network representations designed to overcome such limitations. Performed with an unprecedented level of accuracy in geometric approximation at microscale, the pore scale flow simulations predict experimental permeabilities measured at lab scale in the same rock sample without the need for calibration or correction. By applying the method to a broader class of representative geological samples, with permeability values covering two orders of magnitude, we obtain scaling relationships that reveal how mesoscale permeability emerges from microscopic capillary diameter and fluid velocity distributions.

## Introduction

Permeability is a critical figure-of-merit in the characterization of porous geological samples for applications ranging from water management to fluid recovery and carbon dioxide sequestration [1–8]. Fluid flow assessment in rocks typically involves multiple physical length scales; a fluid passes through a complex, interconnected network of capillaries with diameters ranging from nanometers to millimeters, while flow conditions are typically set and measured at lab scale, see Figure 1. Once the spatial distribution of the connected pore space in a rock sample is known, a flow model is applied to computationally predict the fluid permeability based on the capillary network geometric boundaries [9–14]. We show in the following that a high-accuracy, network-based representation of a microscopic fraction of a rock's connected pore space is a suitable template for computationally predicting experimental permeability results obtained from the same rock sample at lab scale, a volume upscaling by 3 orders of magnitude. An application of the method to 11 sandstone samples, see Table 1, reveals permeability scaling as function of diameter distribution and flow speed. We suggest that the extracted slopes be used more generally for characterizing permeability scaling in this geological sample class.

Methodically, the application of X-ray microtomography to rock samples provides a series of images of the spatial distribution of the pore space from which three-dimensional, digital rock representations are created [15–17]. Once a full series of microscopic rock images is acquired, the image series undergoes a sequence of processing steps with regards to noise, segmentation and morphology, producing a data cube (digital rock) containing voxels that either represent solid or void space of the imaged rock [18]. Here, the spatial discretization method – such as meshes and grids, used by Finite-Element, Finite-Volume and Lattice Boltzmann; or nodes and edges, used by network-based methods – depends on the flow simulation algorithm to be used for performing subsequent permeability predictions [19–24].

*Table 1: List of sandstone rock samples used in our study and their respective properties. Sample labels A-K are listed in increasing order of permeability. We estimate the experimental error to be ±0.5% for porosity and ±10% for permeability, respectively [35].*

| Sample | Name | Porosity (%) | Permeability (mD) |
|---|---|---|---|
| A | Bandera Gray | 18.10 | 9 |
| B | Parker | 14.77 | 10 |
| C | Kirby | 19.95 | 62 |
| D | Bandera Brown | 24.11 | 63 |
| E | Berea Sister Gray | 19.07 | 80 |
| F | Berea Upper Gray | 18.56 | 86 |
| G | Berea | 18.96 | 121 |
| H | Castlegate | 26.54 | 269 |
| I | Buff Berea | 24.02 | 275 |
| J | Leopard | 20.22 | 327 |
| K | Bentheimer | 22.64 | 386 |

In any of these methods, the tradeoff between the voxel (volume pixel) size and the total imaged sample volume poses practical limits to the spatial resolution of lab scale samples in which permeability measurements are typically performed. The computational representations of lab scale rock sample do not fully resolve the diameter distribution of the capillary network, which consequently leads to inaccuracies in permeability predictions. In addition, computational approximations of the connected pore space by geometrical primitives, such as balls and sticks, can be insufficient to capture the actual complexity of the capillary network at pore scale. Sample heterogeneities can further complicate the picture; the higher the spatial heterogeneity of a rock sample the larger the Representative Elementary Volume (REV) for quantifying the threshold sampling volume at which the statistical properties of the capillary network become representative for the entire lab scale rock sample [25–28]. In essence, a combination of stringent requirements needs to be met for successfully predicting the permeability of lab scale samples based on microscopic capillary network representations.

For establishing those requirements, we have systematically analyzed a representative group of highly resolved, three-dimensional microscopic image cubes measured on rock samples for which permeability and porosity were experimentally verified at lab scale. For evaluating the influence of pore scale representation quality on the accuracy of permeability predictions, we have used three candidate network representations as geometrical template: (1) the *Capillary Network Model* (CNM), which transforms the pore space into a voxel-wide line at the center of the pore channels; (2) the *Reduced Max Ball Model* (RBM) [29], in which the network is constructed using connecting cylinders modelled by longer line segments that follow the medial axis of the pore space; and (3) the *Pore Network Model* (PNM) [30], which divides pore space into pores and throats, where each pore is a node in the network and each throat is a link between nodes. For each sample, we have compared the computational permeability predictions obtained for each of the three pore space representations with the experimental permeability obtained from the same rock sample at lab scale. By aggregating the data of all samples studied, we have obtained the scaling relationships and slopes for characterizing more broadly the permeability of the entire sandstone sample class.

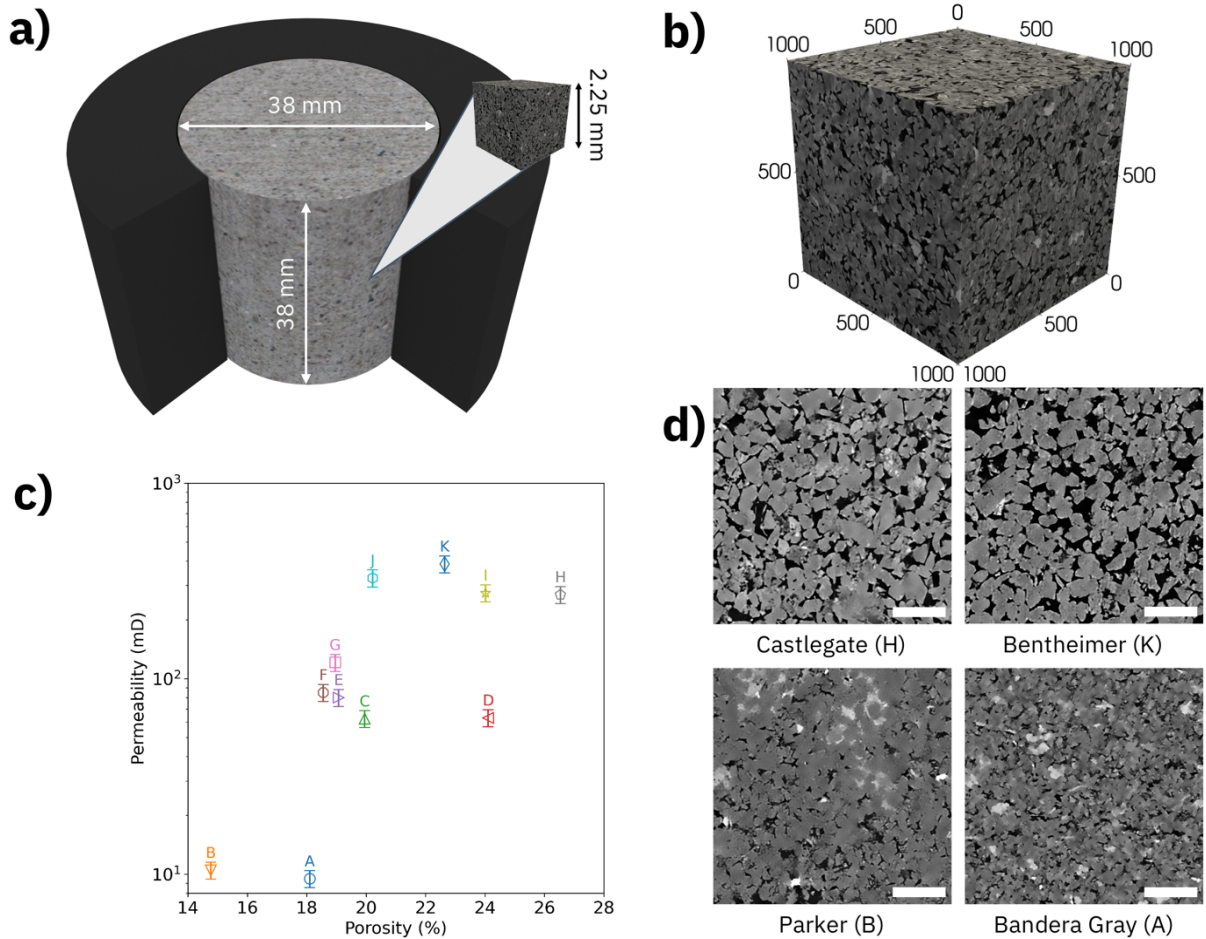

*Figure 1: Conception of permeability and porosity determination in rock samples. (a) Permeability and porosity measurements are performed at lab scale. However, computational simulations are performed with microscopic network representation at pore scale. (b) Pore scale representation of a rock sample derived from an X-ray microtomography with 1000 voxels of 2.25 µm along each side, resulting in an overall side length of 2.25 mm. (c) The experimental porosity-permeability plot of containing all rock samples. (d) Representative grayscale images extracted from the data cube of the least porous (B), least permeable (A), most porous (H) and most permeable (K) rock sample, respectively, exhibit the microstructural variation occurring at pore scale. The white scale bar at the lower right corner of each image represents 500 µm.*

## Results

*Experimental characterization*

We have measured the porosity and permeability of a set of 11 cylindrical sandstone samples with radii and heights of 19 and 38 mm, respectively. The connected porosities found in these samples range from 14% to 27%, with a mean of 20±3%, and the absolute permeabilities from 9mD to 386 mD, with a mean of 150±40 mD, as summarized in Table 1. Details on the flow measurements are provided in the Methods Section. In Figure 1(c), we plot the measured gas permeability and porosity for each sample. The microscopic distributions of connected pore space in each rock are mainly responsible for the observed porosity/permeability variations as can be examined by visual inspection of the X-ray microtomography images of the samples. In Figure 1(d), we show representative images taken at the center of the image cube of the least porous (B), least permeable (A), most porous (H) and most permeable (K) rock samples in the dataset.

We have acquired X-ray microtomography image cubes on sub-samples (height = 30 mm, radius = 5 mm) extracted from each of the rock plugs listed in Table 1 (see Methods Section for details). The representative three-dimensional microtomography in Figure 1(b) has a voxel size of 2.25 μm in which the lighter (darker) gray levels correspond to solid (void) space in the sandstone. For extracting the connected pore space, we have processed each raw image (gray scale) cube through a workflow that includes contrast-enhancement, noise reduction and threshold-based segmentation (see Methods Section for details). The resulting binary image cube contains the spatial map of pore space inside the sample. The volume fraction of voxels identified as pores is readily available, however, a subsequent processing step is needed to remove isolated pores that do not contribute to permeability.

We have then used the connected pore space in the image cube as a template for performing numerical flow simulations and permeability predictions. By analyzing the dependence of flow simulation accuracy on rock image cube size, see Supplementary Figure S2, we obtain REV = (2.25 mm)$^3$. Finally, we obtain a volume scaling factor of $V_{exp}$/REV = 3784, which relates the connected pore space volume used for computing permeability predictions to the reference sample volume that was probed in our lab experiments.

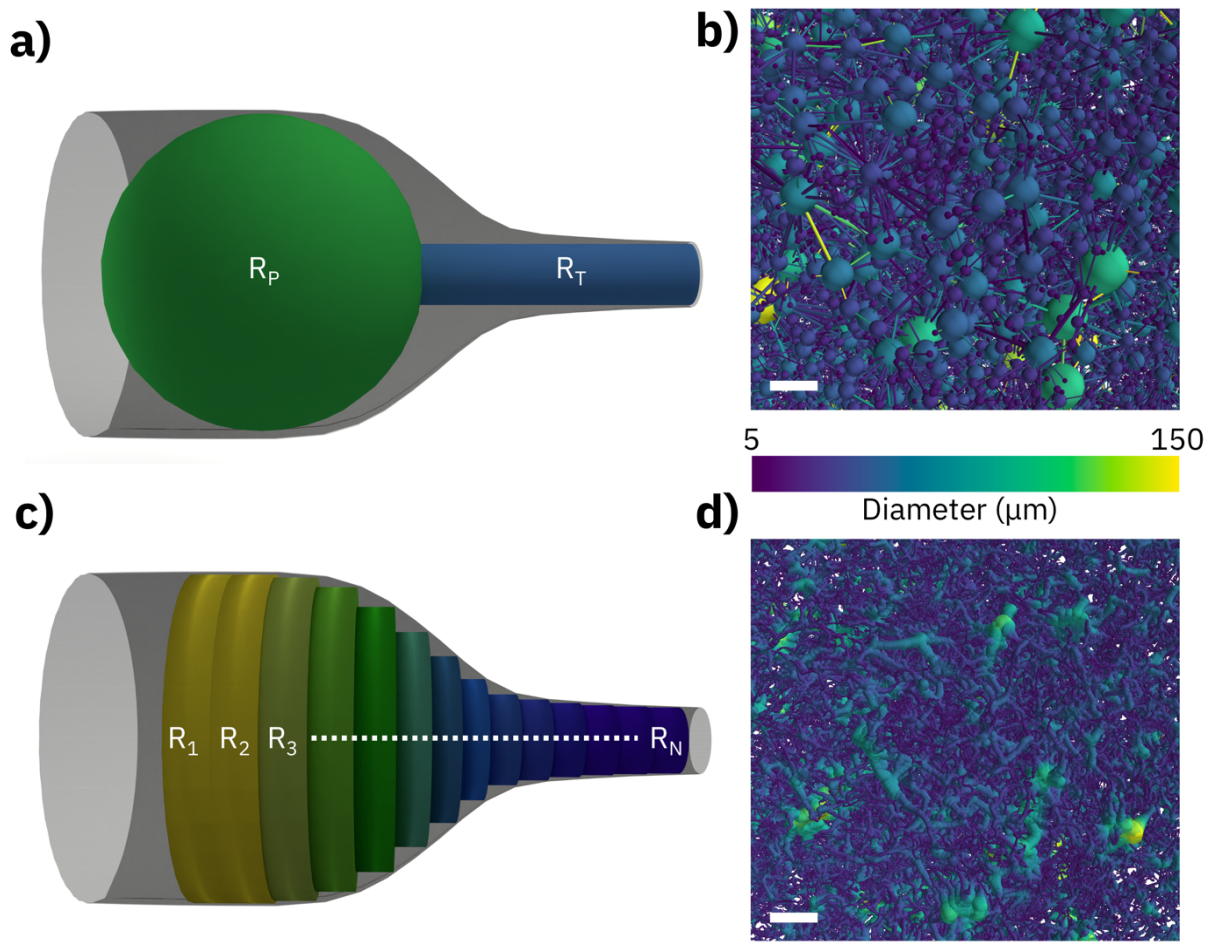

*Figure 2: Visual conceptions (left) and implementation examples (right) of representative network models. (a) The Pore Network Model separates the pore space into spherical pores with radius $R_P$ and cylindrical throats with radius $R_T$. (b) Pore Network Model representation using color-coded diameters. (c) The capillary network model separates the pore space into a sequence of short cylinders with gradually changing radii $\{R_n\}$. (d) Capillary Network Model representation using color-coded diameters. The white scale bar represents 100 μm.*

*Numerical simulations*

For quantifying the accuracy of geometric approximation in the digital rocks, we have compared three network-based representations of the microscopic pore space – PNM as algorithmically implemented in the PoreSpy open-source package [31], RMB based on the algorithm developed by Andreeta *et al.* [29] and CNM performed by an algorithm reported in this paper for the first time.

In Figure 2, we show pore network (top) and capillary network (bottom) representations applied to the same sample region. In a pore network representation, see Figure 2(a), the void space is subdivided into pores and throats. Pore and throat are each represented by their radius, $R_P$ and $R_T$, respectively, while specifics of the cross-sectional geometries are matched using shape factors [14]. In Figure 2(b), a network of spheres and cylinders geometrically approximates the occurring void space. For comparison, Figure 2(c) exemplifies how a capillary network representation would approximate the same sample region. Here, the void space is filled with short (one voxel long) cylinders having their radii $\{R_n\}$ evolving gradually to match the local boundaries. The resulting network of connected capillaries in this approximation is shown in Figure 2(d).

In PNM, nodes are associated with pores, which have a finite volume, and links are associated with throats, which impose a resistance to the flow between the two nodes it connects. In CNM, on the other hand, nodes are considered zero-volume points that do not contribute to the pore space, while links represent finite-volume cylinders that account for all the pore volume and flow resistance. Finally, RMB has features of both pore and capillary network representations as it uses extended cylinders to represent the pore space.

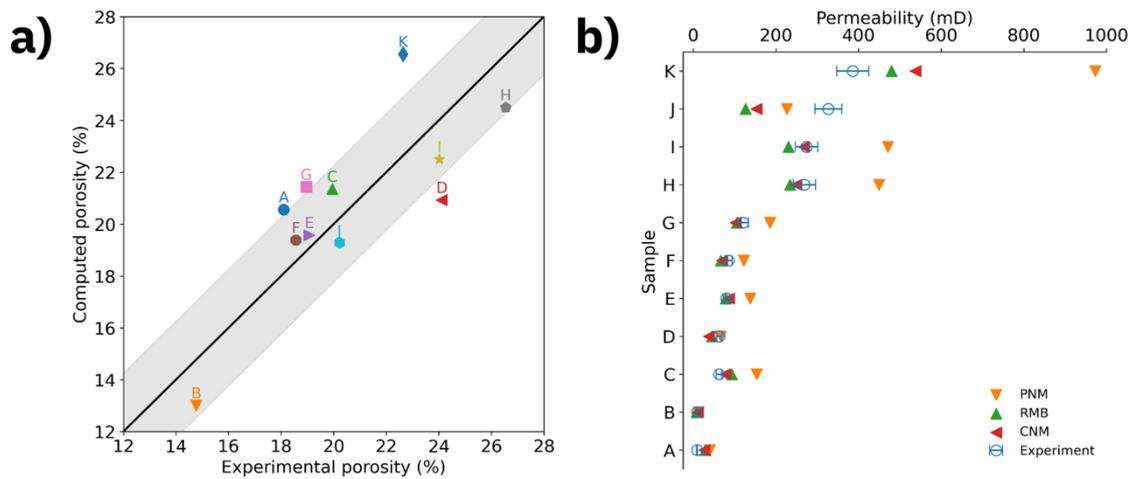

*Figure 3: Comparison between experimental and computed porosities and permeabilities. (a) Computed vs. experimental porosity for all samples studied. The solid line indicates agreement between computation and experiments and the gray shaded area represents the root mean squared error of 2.25 percentage points. (b) Experimental and simulated permeability for all samples studied. The computational results represent the mean permeability along the three main axes. Experimental results are represented by (blue) open circles, those from the Pore Network Model by (orange) filled down-pointing triangles, those from the Reduced Max Ball Model by (green) filled up-pointing triangles and those from the Capillary Network Model by (red) filled left-pointing triangles.*

In Figure 3, we have compared the experimental and computational porosities and permeabilities. The computational porosity values extracted from the connected pore space in digital rocks are plotted against the experimental porosity values in Figure 3(a). The diagonal indicating a perfect match between experiment and computational prediction is overlaid by a gray shaded area indicating a mean squared error of 2.25%. Nine of eleven samples fall within the shaded area, indicating reasonable agreement.

In Figure 3(b), we plot the experimental and simulated permeabilities for each network model considered in this study. In each of the three network representations, we have simulated water flow assuming a 10 kPa/m pressure gradient across the sample. The flow simulation algorithm solves a linear system of equations that applies Poiseuille law at each network link (i.e., cylindrical capillary) and mass conservation law at each network node (for details see Method Section). The simulated permeabilities represent the quadratic mean of the permeability along the three main axes. Considering the entire sandstone sample set, we observe the best agreement for CNM with a mean relative error between model prediction and experimental result of 38%, followed by RMB with 42%, attesting to the high accuracy of the geometrical approximation achieved with this approach. In contrast, the PNM based predictions are significantly less accurate, at a mean relative error of 92%, with a maximum mismatch of more than a factor of two in the case of sample K.

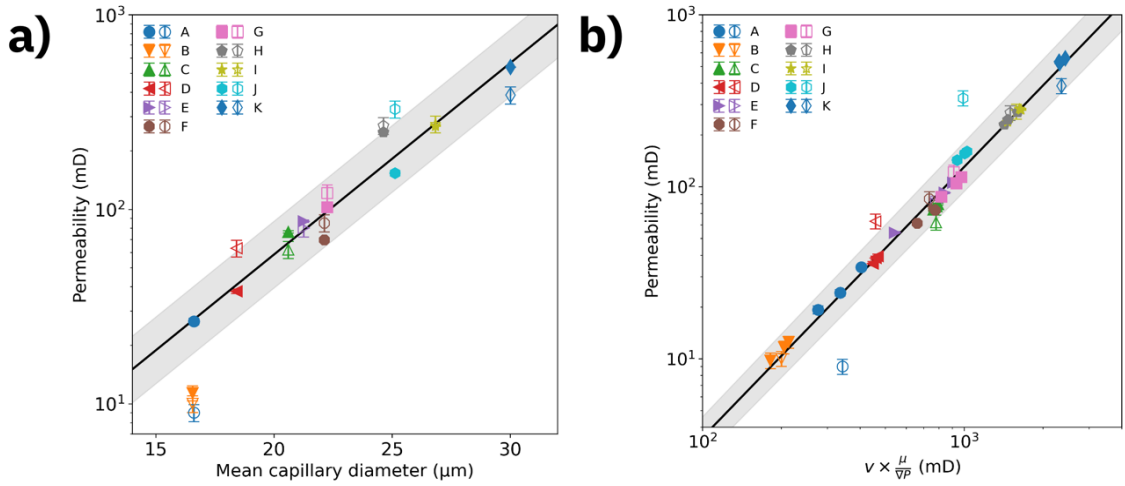

*Figure 4: Scaling of permeability as function of mean capillary diameters and mean flow speeds. Computed (experimental) permeability is represented by filled (open) symbols. (a) Permeability as function of mean capillary diameter for all rock samples studied. (b) Permeability as function of volume-averaged flow speed multiplied by ($\mu/\nabla P$) along all 3 axes for all rock samples studied. The plot aggregates results from simulation scenarios with viscosity $\mu$=1 cP and pressure gradient of 10 kPa/m, as well as variations with 10x higher viscosity or 10x stronger gradient, covering two orders of magnitude variation in flow speeds. The lines represent linear fits to the data and the shaded area represents the fit uncertainty.*

As a key result of this investigation, we plot in Figure 4, for each sample, the relationship between lab scale permeability and microscopic properties based on CNM. While bulk permeability and porosity of the sample class are broadly scattered, see Figure 1, the permeability (both experimental and computed) scales with the mean capillary diameter observed in the samples, see Figure 4(a). A linear fit to the data provides a slope of $(0.098 \pm 0.007)\langle D\rangle[\mu m]$ for $\log_{10} K[mD]$, which characterizes the entire sandstone sample set. In Figure 4(b), we plot experimental and computed permeability as a function of generalized velocity $u = (v \times \mu/\nabla P)$, where $v$ is the (microscopic) volume-weighted average flow speed inside the capillaries, $\mu$ is the dynamic viscosity of the fluid and $\nabla P$ is the pressure gradient along the flow axis. A linear fit to the data ($R^2 > 0.98$) in a $\log_{10} K[mD] \times \log_{10} u[mD]$ plot of indicates that permeability is well-described by $(0.0024\pm0.0003) \times u^{1.57\pm0.02}$ for the entire set of sandstone samples.

A practical application of the scaling results is estimating lab scale permeabilities based on mean capillary diameters and flow speeds taken from CNM predictions performed on X-ray microtomography data or, conversely, getting insight into microscopic flow speeds based on lab scale permeability measurements.

## Discussion

We have computationally predicted lab scale permeabilities based on X-ray microtomography data for a sample set containing 11 sandstones. A high-accuracy, network-based geometrical representation of a microscopic volume fraction of the connected pore space of a rock was used as a basis for performing fluid flow simulations. This representation enables predicting experimental permeability results obtained from the same rock sample at lab scale (more that 3000x volume) with a mean relative error between microscopic model prediction and lab scale experimental result of 38%. By aggregating the results, we have obtained scaling relationships between the lab scale permeabilities and the mean capillary diameter and flow speed, respectively, for the sandstone sample class that can be used for estimating lab scale permeabilities based on X-ray microtomography data. Future research should extend the above methodology to geological sample classes with higher degrees of heterogeneity and more complex capillary diameter distributions, such as carbonate and shale rocks.

## Materials and Methods

### Rock samples

We have acquired a set of sandstone rocks (*Kocurek Industries INC.*) with the following 11 samples: Bandera Gray (A), Parker (B), Kirby (C), Bandera Brown (D), Berea Sister Gray (E), Berea Upper Gray (F), Berea (G), Castlegate (H), Buff Berea (I), Leopard (J) and Bentheimer (K).

### Experimental characterization

We have experimentally characterized the cylindrical plug samples (height = 38mm, radius = 19mm) at lab scale following the API RP-40 norm [32]. In a first step, we have measured porosity through Helium pressure variations in a Boyle's Law Double Cell. In a second step, we have measured permeability by monitoring steady-state Nitrogen flow rate in the axial direction of the cylindrical samples [33]. An axial pressure gradient of 30-40 psi was applied to the samples for establishing laminar flow conditions. For restricting fluid flow to the axial direction, the plugs were confined laterally by applying an axial pressure through a rubber sealing surrounding the sample. The radial pressure was limited to 500 psi so to avoid significant modifications of the plug's pore volume. A Klinkenberg correction [34] was applied to correct for fluid expansion. We estimate the experimental error to be ±0.5% for porosity and ±10% for permeability, respectively [35].

### X-ray microtomography

In a next step, the plugs were sub-sampled to obtain smaller cylindrical samples (height = 30 mm, radius = 5 mm). We have obtained pore scale image data by using high-resolution 3D X-ray Microtomography (SkyScan 1272, Bruker). Source voltage and current were set to 50 kV and 200 µA, respectively. The CCD camera was configured to acquire projections of 4904 x 3280 pixels, resulting in a pixel side length of 2.25 µm. We have performed image reconstruction using SkyScan NRecon (version: 1.7.0.4, Bruker).

### Image processing

We have processed the rock image data into subsets containing $1000^3$ cubic voxels and converted the image from 16-bit to 8-bit gray scale. We have then applied an enhancement filter to equalize the contrast across multiple images. For each data set, we have cut off the grayscale level where the accumulated grayscale histogram achieved 99.8% and mapped the remaining grayscale levels to the [0, 255] interval. To reduce image noise, we have executed on each data set a 3D non-local means filter [36–37] available in Fiji [38] using a smoothing factor of 1 and automatically estimated sigma [39] parameters. Finally, by using a threshold level calculated by the IsoData method [40], we have segmented the noise-reduced grayscale images into solid and void space leading to a binary image. The image-processing parameters for each sample are available in Supplementary Table S1. We have then processed the binary images using the Enhanced Hoshen-Kopelman algorithm [41] for morphological analysis and, in a final step, eliminated from each data cube the pore voxels that are not connected to the percolating network.

*Network extraction*

We have then used the binary image data sets containing the connected pore voxels as input to three representative network extraction algorithms: PNM, RMB and CNM. Briefly, all algorithms calculate distance maps of the images and construct a hierarchical graph of the voxels associated with the void space.

The PNM is extracted using the SNOW algorithm based on watershed transformation [30], available in PoreSpy [42] version 1.2.0, which is specialized in extracting pore networks from high-porosity materials.

The RMB algorithm [29] is a modified medial-axis extraction based on the original Max Ball Algorithm. The pore-throat connections in the Max Ball Algorithm are found through the maximum ball chains. These chains are further processed in the RMB by finding the optimum fluid flow paths between pore centers through Dijkstra's algorithm. The final output is a simplified medial axis composed by the spheres on the optimum paths' chains. The sphere´s centers are the nodes in the network and the links are modelled by the neighboring spheres radii and distance between centers.

The CNM representation is based on the *Centerline* algorithm [43]. A centerline is a thin, one-dimensional object that captures a 3D object's main symmetry axes, summarizing its main shape into a set of curves [44]. In this work, we have used for the first time an algorithm that extracts the centerlines of a 3D image by using an adaptation of the traditional Dijkstra's Minimum Path algorithm in a graph with penalized distance. The full algorithm description can be found in the Supplementary Information.

*Fluid flow simulation*

The flow simulation algorithm applied in this study operates on nodes and links of the network representations. Specifically, the algorithm applies Poiseuille law to the links and mass conservation law to the interior nodes, while maintaining a fixed pressure difference between inlet and outlet boundary nodes. This is represented by a system of mass conservation equations $\Sigma_j Q_{ij} = 0$ for all nodes i, where $Q_{ij} = (\pi R_{ij}^4/8\mu L_{ij}) (P_i-P_j)$ is the flow rate in the capillary that connects node i to node j. The geometrical parameters R and L, respectively, represent the radius and the length of a capillary (link) connecting two nodes of the network and µ is the dynamic viscosity of the fluid. Unless stated otherwise, all simulations used a viscosity of µ = 1 cP and applied a 10 kPa/m pressure gradient along the flow direction in addition to atmospheric pressure. In order to validate the flow simulations, we have benchmarked them against OpenPNM [45] version 2.4.2, with results matching within ±1 mD.

## Data availability

The microtomography datasets generated and/or analyzed during the current study are available in the Digital Rocks Portal repository, at https://dx.doi.org/10.17612/f4h1-w124.

## Code availability

The code used to obtain the PNM results reported in the current study are available in a Github repository, at https://github.com/mbandreeta/absolute_permeability_simulations.

## References

1. Su, E., Liang, Y., Zou, Q., Niu, F. & Li, L. Analysis of Effects of $CO_2$ Injection on Coalbed Permeability: Implications for Coal Seam $CO_2$ Sequestration. *Energy and Fuels* **33**, 6606–6615 (2019).
2. Niu, Q. & Zhang, C. Permeability Prediction in Rocks Experiencing Mineral Precipitation and Dissolution: A Numerical Study. *Water Resour. Res.* **55**, 3107–3121 (2019).
3. Zhang, P., Celia, M. A., Bandilla, K. W., Hu, L. & Meegoda, J. N. A Pore-Network Simulation Model of Dynamic $CO_2$ Migration in Organic-Rich Shale Formations. *Transp. Porous Media* **133**, 479–496 (2020).
4. Al-Janabi, A. M. S., Yusuf, B. & Ghazali, A. H. Modeling the Infiltration Capacity of Permeable Stormwater Channels with a Check Dam System. *Water Resour. Manag.* **33**, 2453–2470 (2019).
5. Millington, R. & Quirk, J. P. Permeability of porous solids. *Trans. Faraday Soc.* **57**, 1200–1207 (1961).
6. Yu, X., Hong, C., Peng, G. & Lu, S. Response of pore structures to long-term fertilization by a combination of synchrotron radiation X-ray microcomputed tomography and a pore network model. *Eur. J. Soil Sci.* **69**, 290–302 (2018).
7. Roth, K. Fluids in Porous Media. *Soil Phys.* **3**, 39–67 (2012).
8. Berg, C. F. & Held, R. Fundamental Transport Property Relations in Porous Media Incorporating Detailed Pore Structure Description. *Transp. Porous Media* **112**, 467–487 (2016).
9. Zheng, D. & Reza, Z. Pore-network extraction algorithm for shale accounting for geometry-effect. *J. Pet. Sci. Eng.* **176**, 74–84 (2019).
10. Liang, Y., Hu, P., Wang, S., Song, S. & Jiang, S. Medial axis extraction algorithm specializing in porous media. *Powder Technol.* **343**, 512–520 (2019).
11. Rabbani, A., Mostaghimi, P. & Armstrong, R. T. Pore network extraction using geometrical domain decomposition. *Adv. Water Resour.* **123**, 70–83 (2019).
12. Silin, D. & Patzek, T. Pore space morphology analysis using maximal inscribed spheres. *Phys. A Stat. Mech. its Appl.* **371**, 336–360 (2006).
13. Arand, F. & Hesser, J. Accurate and efficient maximal ball algorithm for pore network extraction. *Comput. Geosci.* **101**, 28–37 (2017).
14. Raeini, A. Q., Bijeljic, B. & Blunt, M. J. Generalized network modeling: Network extraction as a coarse-scale discretization of the void space of porous media. *Phys. Rev. E* **96**, 1–17 (2017).
15. Saxena, N., Hows, A., Hofmann, R., Freeman, J. & Appel, M. Estimating Pore Volume of Rocks from Pore-Scale Imaging. *Transp. Porous Media* **129**, 403–412 (2019).
16. Saxena, N. et al. Effect of image segmentation & voxel size on micro-CT computed effective transport & elastic properties. *Mar. Pet. Geol.* **86**, 972–990 (2017).
17. Cnudde, V. & Boone, M. N. High-resolution X-ray computed tomography in geosciences: A review of the current technology and applications. *Earth-Science Rev.* **123**, 1–17 (2013).
18. Andrä, H. et al. Digital rock physics benchmarks—Part I: Imaging and segmentation. *Comput. Geosci.* **50**, 25–32 (2013).

19. Blunt, M. J. et al. Pore-scale imaging and modelling. *Adv. Water Resour.* **51**, 197–216 (2013).
20. Bernabé, Y., Li, M., Tang, Y.-B. & Evans, B. Pore space connectivity and the transport properties of rocks. *Oil Gas Sci. Technol. – Rev. d'IFP Energies Nouv.* **71**, 50 (2016).
21. Tang, Y. B. et al. Pore-scale heterogeneity, flow channeling and permeability: Network simulation and comparison to experimental data. *Phys. A Stat. Mech. its Appl.* **535**, (2019).
22. Bultreys, T., De Boever, W. & Cnudde, V. Imaging and image-based fluid transport modeling at the pore scale in geological materials: A practical introduction to the current state-of-the-art. *Earth-Science Rev.* **155**, 93–128 (2016).
23. Aghaei, A. & Piri, M. Direct pore-to-core up-scaling of displacement processes: Dynamic pore network modeling and experimentation. *J. Hydrol.* **522**, 488–509 (2015).
24. Saxena, N. et al. References and benchmarks for pore-scale flow simulated using micro-CT images of porous media and digital rocks. *Adv. Water Resour.* **109**, 211–235 (2017).
25. Miao, X., Gerke, K. M. & Sizonenko, T. O. A new way to parameterize hydraulic conductance's of pore elements: A step towards creating pore-networks without pore shape simplifications. *Adv. Water Resour.* **105**, 162–172 (2017).
26. Bondino, I., Hamon, G., Kallel, W. & Kac, D. Relative Permeabilities From Simulation in 3D Rock Models and Equivalent Pore Networks: Critical Review and Way Forward1. *Petrophysics* **54**, 538–546 (2013).
27. Baychev, T. G. et al. Reliability of Algorithms Interpreting Topological and Geometric Properties of Porous Media for Pore Network Modelling. *Transport in Porous Media* **128**, 271-301 (2019).
28. Gerke, K. M. et al. Improving watershed-based pore-network extraction method using maximum inscribed ball pore-body positioning. *Adv. Water Resour.* **140**, 103576 (2020).
29. Barsi-Andreeta, M., Lucas-Oliveira, E., de Araujo-Ferreira, A. G., Trevizan, W. A. & Bonagamba, T. J. Pore Network and Medial Axis simultaneous extraction through Maximal Ball Algorithm. Preprint at https://arxiv.org/abs/1912.04759 (2019).
30. Gostick, J. T. Versatile and efficient pore network extraction method using marker-based watershed segmentation. *Phys. Rev. E* **96**, 1–15 (2017).
31. Gostick, J. et al. PoreSpy: A Python Toolkit for Quantitative Analysis of Porous Media Images. *J. Open Source Softw.* **4**, 1296 (2019).
32. API. RP40 in *Recommended practices for core analysis* (American Petroleum Institute, 1998).
33. Tanikawa, W. & Shimamoto, T. Klinkenberg effect for gas permeability and its comparison to water permeability for porous sedimentary rocks. *Hydrol. Earth Syst. Sci. Discuss.* **3**, 1315–1338 (2006).
34. Klinkenberg, L. J. The Permeability of Porous Media to Liquids and Gases in *Drill. Prod. Pract.* (American Petroleum Institute, 1941).
35. Thomas, D. & Pugh, V. A statistical analysis of the accuracy and reproducibility of standard core analysis. *Petrophysics* **30**, 71–77 (1989).
36. Buades, A., Coll, B. & Morel, J.-M. Non-Local Means Denoising. *Image Process. Online* **1**, 208–212 (2011).
37. J. Darbon, A. Cunha, T. F. Chan, S. Osher and G. J. Jensen, Fast Nonlocal Filtering Applied To Electron Cryomicroscopy, *5th IEEE International Symposium on Biomedical Imaging: From Nano to Macro*, Paris, pp. 1331-1334 (2008).
38. Schindelin, J. et al. Fiji: an open-source platform for biological-image analysis. *Nat. Methods* **9**, 676–682 (2012).
39. Immerkær, J. Fast Noise Variance Estimation. *Comput. Vis. Image Underst.* **64**, 300–302 (1996).

40. Picture Thresholding Using an Iterative Selection Method. *IEEE Trans. Syst. Man. Cybern.* **8**, 630–632 (1978).
41. Hoshen, J., Berry, M. W. & Minser, K. S. Percolation and cluster structure parameters: The enhanced Hoshen-Kopelman algorithm. *Phys. Rev. E* **56**, 1455–1460 (1997).
42. GOSTICK, Jeff T. et al. PoreSpy: A Python Toolkit for Quantitative Analysis of Porous Media Images. *Journal of Open Source Software* **4**, 1296 (2019).
43. Telea, A. & Vilanova, A. A Robust Level-Set Algorithm for Centerline Extraction. In *Proceedings of the Symposium on Data Visualisation,* 185–194 (Eurographics Association, 2003).
44. Niblack, C. W., Capson, D. W. & Gibbons, P. B. Generating Skeletons and Centerlines from The Medial Axis Transform. In *Proceedings of 10th International Conference on Pattern Recognition* vol. I, 881–885 (1990).
45. Gostick, J. et al. OpenPNM: A Pore Network Modeling Package. *Comput. Sci. Eng.* **18**, 60–74 (2016).
46. Bitter, I., Kaufman, A. E. & Sato, M. Penalized-Distance Volumetric Skeleton Algorithm. *IEEE Trans. Vis. Comput. Graph.* **7**, 195–206 (2001).
47. Zhou, Y., Kaufman, A. & Toga, A. W. Three-Dimensional Skeleton and Centerline Generation Based on an Approximate Minimum Distance Field. *Vis. Comput.* **14**, 303–314 (1998).
48. Tarjan, R. Depth-First Search and Linear Graph Algorithms. *SIAM J. Comput.* **1**, 146–160 (1972).
49. Dijkstra, E. W. A Note on Two Problems in Connexion With Graphs. *Numer. Math.* **1**, 269–271 (1959).

## Acknowledgements

E.L.O., M.B.A. and T.J.B. acknowledge National Council for Scientific and Technological Development (CNPq), grants (140215/2015-8), (153627/2012-3) and (308076/2018-4), respectively. M.B.A. and T.J.B. acknowledge Centro de Pesquisa e Desenvolvimento Leopoldo Américo Miguez de Mello (CENPES/Petrobras) grants (2014/00389-8, 2015/00416-8) and (2014/00389-8), respectively. R.F.N. and M.S. acknowledge Alexandre Ashade, Peter Bryant, William Candela, Ítalo Nievinski, Giulia Coutinho (IBM Research) for their contributions to earlier versions of the CNM algorithm and Ricardo Ohta (IBM Research) for help with the 3D rendered images. Authors would like to thank T.J.B. (USP) and Ulisses Mello (IBM Research) for making the collaboration between these institutions possible.

## Author contributions

M.S. and T.J.B. conceived the work. W.A.T. performed porosity and permeability measurements at lab scale. E.L.O. acquired X-ray microtomography images. H.B. developed the capillary network extraction algorithm. R.F.N. and M.B.A. performed the computer simulations and data analysis. R.F.N. and M.S. wrote the paper with input from all the co-authors.

## Additional Information

*Competing interests*
The authors declare no competing interests.

# Supplementary Information

*Image-processing method parameters*

*Supplementary Table S1: Image-processing parameters for rock image cubes. Prior to segmentation, the contrast enhancement filter cut off the grayscale histogram at the described cut-off level and rescaled all grayscale levels to [0, 255]. Finally, after running the non-local means filter, the grayscale image was segmented using a threshold level calculated by the IsoData algorithm.*

| Sample | Name | Cut-off level | Threshold level |
|---|---|---|---|
| A | Bandera Gray | 93 | 81 |
| B | Parker | 63 | 72 |
| C | Kirby | 61 | 75 |
| D | Bandera Brown | 73 | 80 |
| E | Berea Sister Gray | 90 | 54 |
| F | Berea Upper Gray | 101 | 48 |
| G | Berea | 71 | 59 |
| H | Castlegate | 156 | 81 |
| I | Buff Berea | 62 | 73 |
| J | Leopard | 76 | 72 |
| K | Bentheimer | 200 | 71 |

*Computed porosity and permeability*

*Supplementary Table S2: Experimental and computed results for porosity and permeability. Porosity results compare experimental measurements to values computed from microtomography data. Permeability results compare experimental values to Pore Network Model (PNM), Reduced Max Ball Model (RMB) and Capillary Network Model (CNM) values. We estimate the experimental error to be ±0.5% for porosity and ±10% for permeability, respectively [35].*

| Sample | Name | Porosity (%) | | Permeability (mD) | | | |
|---|---|---|---|---|---|---|---|
| | | Exp. | μCT | Exp. | PNM | RMB | CNM |
| A | Bandera Gray | 18.10 | 20.56 | 9 | 40 | 29 | 26 |
| B | Parker | 14.77 | 13.00 | 10 | 13 | 8 | 11 |
| C | Kirby | 19.95 | 21.35 | 62 | 154 | 93 | 76 |
| D | Bandera Brown | 24.11 | 20.93 | 63 | 67 | 45 | 38 |
| E | Berea Sister Gray | 19.07 | 19.57 | 80 | 138 | 79 | 86 |
| F | Berea Upper Gray | 18.56 | 19.38 | 85 | 122 | 66 | 70 |
| G | Berea | 18.96 | 21.43 | 121 | 186 | 106 | 102 |
| H | Castlegate | 26.54 | 24.50 | 269 | 450 | 234 | 249 |
| I | Buff Berea | 24.02 | 22.50 | 274 | 471 | 230 | 267 |
| J | Leopard | 20.22 | 19.28 | 327 | 226 | 126 | 153 |
| K | Bentheimer | 22.64 | 26.56 | 386 | 973 | 480 | 538 |

*Capillary network extraction algorithm*

A commonly used formal definition of the centerline, which is agnostic to the extraction algorithm, requires these four properties to be obeyed:

1. **Connected:** There must have at least one path between any two voxels of the centerline.
2. **Centered**: Any voxel of the centerline must be centered with respect to the object's boundary.
3. **Thin**: Centerline should be only 1-voxel thick.

4. **Insensitive to boundary noise:** Small surface details should not produce large twists or numerous small branches in the centerline.

In the following, we outline a novel adaptation of the Dijkstra shortest path algorithm to extract the centerline, or CNM, from a rock tomography image cube. The algorithm uses a penalization function and gradient vector for building a centerline with minimal centerline property violation, while detecting and connecting image voxels that form cycles in the image.

In a first step, the algorithm establishes the distance map of the object by computing for each pore space voxel its closest distance to the boundary. The distance is determined by the recursive function $d_{min}(v)$ that calculates for a given voxel $v$ the distance from $v$ to the closest boundary voxel. The boundaries voxels $v_b$ are set as the base case of the recursion; $d_{min}(v_b) = 0$. For all other voxels $v$, their distance from the closest boundary voxel is defined as $d_{min}(v) = \min_{t \in N(v)} \{d_{min}(t) + E(v,t)\}$, where $N(v)$ is the set of $v$'s neighbor voxels $E(v,t)$ is the Euclidian distance between $v$ and $t$. In the definition of neighborhood applied here, voxels $v$ and $t$ are neighbors if they share a Face, Edge or Vertex, leading to Euclidian distances of 1, $\sqrt{2}$ and $\sqrt{3}$, respectively.

The algorithm uses a priority list as its main data structure for iteratively computing $d_{min}(v)$ for every voxel $v$. First, boundary voxels $v_b$ are added to the list with their priority set to 0. Then, in the main loop, the voxel $t$ with lowest priority (voxel index is used as tiebreaker) is picked from the list and the distance to its neighbor voxel $v$ (that were not yet picked from the list) is set as follow:

- $d_{min}(v) = d_{min}(t) + E(v,t)$ and included into the list, if visited for the first time; or
- $d_{min}(v) = \min\{d_{min}(t) + E(v,t), d_{min}(v)\}$, if already visited.

The process finishes once the priority list is empty. In addition, voxels $v$ for which $d_{min}(v)$ is greater or equal to $d_{min}$ with regards to all adjacent voxels are annotated as local maxima leading to the creation of labeled, local voxel clusters. The main differentiator of our algorithm with regards to published works [43,46,47] is that the source and targets voxels of the centerline to be used in the Dijkstra shortest path algorithm are placed in the facets of the cube that represents the image.

For identifying those voxels, a depth-first search (DFS) algorithm [48] is executed for every voxel residing in the facets of the cube. The DFS algorithm will run multiple times, once for each cluster of connected voxels in the facets. We visualize the outcome of DFS application in Supplementary Figure S1, in which the voxels with the highest values of $d_{min}(v)$ in a cluster (in black) are selected and highlighted (in red).

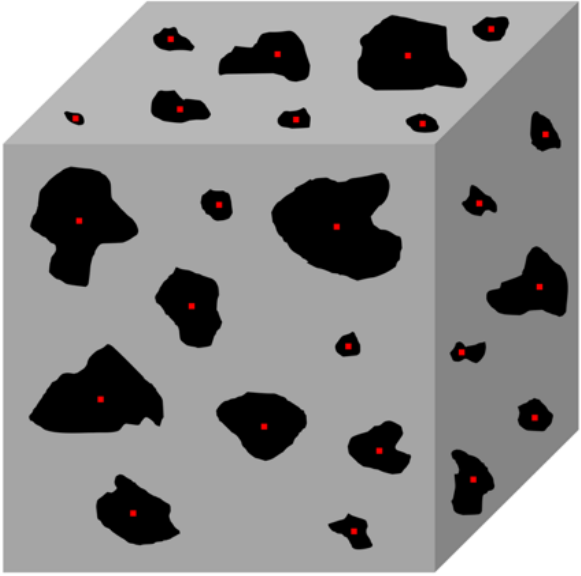

*Supplementary Figure S1: Diagram of the rock sample tomography showing the rock (solid) voxels in gray and porous (void) voxels in black. For each facet, a depth-first search algorithm identifies the most central voxel in each cluster (depicted in red). These voxels are then used as the source and target voxels for the Dijkstra shortest path algorithm.*

Once the process is completed for all facets, one of the selected voxels is set as source voxel and the others are considered target voxels for application of the Dijkstra shortest path algorithm [49]. The output of this algorithm is an acyclic connected graph (tree) consisting of pathways that connect the source voxel to target voxels. The centerline is then built by taking the subset of these pathways that connects source voxel to the target voxels.

In the following, we briefly outline the Dijkstra shortest path algorithm for computing the pathways and the penalization function for minimizing a centerline's property violations. The algorithm relies on two auxiliary data structures: one that stores the cost of shortest path from the source voxel $s$ to each voxel $v$; $c_{min}(v)$; and another one that stores the predecessor voxel of each voxel $v$; $pred(v)$, associated with its respective shortest path to the source voxel. The cost function $c_{min}(v)$ in the original Dijkstra shortest path algorithm is accumulative, that is, it represents the cost of the entire path starting from the source voxel to the target voxel $v$. To compute the centerline, the Dijkstra shortest path algorithm version adapted here only accumulates the penalty cost from the previous step. As a result, the algorithm connects voxel $v$ to the neighbor that better suits the definition of the centerline by being "most centered" rather than the "shortest" path.

Similar to the original version of the algorithm, we use a priority queue to iteratively compute the path from voxel $s$ to each target voxel $v$ that best fits the centerline definition. However, after voxel $t$ is picked up from the priority list, the cost of the path of each of its neighbor voxel $v$ that were not yet picked up from the priority list is updated as follow:

- $c_{min}(v) = penalty(t,v)$ and included into the list, if visited for the first time; or
- $c_{min}(v) = \min\{1 + penalty(pred(t), t) + penalty(t,v) + (1/ d_{min}(v)) * 1E3, c_{min}(v)\}$, if already visited.

In the first case, in which a known path from source to $v$ does not exist, the path cost to traverse the graph from source to voxel $v$ and the predecessor voxel of $v$ are updated ($pred(v) = t$). In the second case, in which a path from the source to $v$ is known, the path cost and the predecessor voxel are updated if, according to the *penalty* function, the connection between voxel $t$ and $v$ better represents the centerline than the existing connection between $pred(v)$ and $v$.

The *penalty* function is vital for the algorithm to heuristically produce pathways matching the centerline definition as close as possible. For avoiding image boundaries, the *penalty* function takes into account a normalized gradient vector computed for every voxel based on distance transformation. The gradient vectors indicate which direction keeps the centerline away from the boundaries. The penalty of going from *t* to *v* depends on the angle α formed by the line that connects voxel *t* to *v* and the gradient vector at voxel *t*; $penalty(t,v) = 0.5 + (sin^2(α)+1)/d_{min}(v)$. However, inside a cluster of voxels annotated as local maximum, we use an alternative *penalty* function that sets a straight pathway.

Finally, we note that typical 3D rock tomography images contain centerlines with cycles. However, the centerline produced by the algorithm described above is acyclic. Therefore, we detect during the centerline algorithm potential pairs of voxels forming centerline cycles. A pair of neighbor voxels *t* and *v* is set to form a potential cycle when the voxel *v* is removed from the queue has its neighbor *t* that was already removed from the queue and satisfies the following conditions: (1) $sin^2(α) < 0.1$, where α is the angle between the line from *t* to *v* and the gradient vector of *v* and, (2), the resultant cycle forms a *LMpath* [47].

*REV determination*

For obtaining the Representative Elementary Volume, we have simulated permeabilities for Berea (G) sample volumes with varying number of voxels in the cube, while keeping the voxel resolution at 2.25 µm. Specifically, we have analyzed 1000 samples with $100^3$ voxels, 512 samples with $125^3$ voxels, 125 samples with $200^3$ voxels, 64 samples with $250^3$ voxels, 8 samples with $500^3$ voxels, and 5 samples with $1000^3$ voxels. In Supplementary Figure S2, we show the resulting simulated permeability values as function of cube side length. We observe that the simulated permeabilities converge towards the experimental value. For a sample size of 1000 voxels (L = 2250 µm), the mean simulated permeability value matches the experimental permeability.

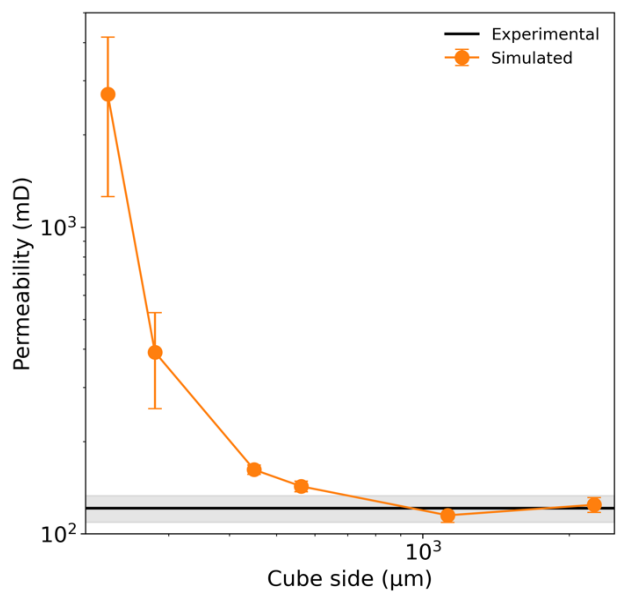

*Supplementary Figure S2: Mean simulated permeability for Berea (G) as function of sample size. Error bars represent the standard error of the mean. The horizontal line indicates the measured permeability and the shaded area the experimental uncertainty. A match is achieved at a cube size of 2250 µm.*

*Distribution of capillary diameters and microscopic flow speeds*

We have investigated the capillary diameter distributions of two representative samples in the set. Supplementary Figures S3(a) and S3(c) show the Parker (B) and Bentheimer (K) samples, respectively. Both Parker and Bentheimer exhibit a peak at around 10-15 µm in their capillary diameter distributions. However, Bentheimer distribution is broader with higher mean diameter of 30 µm, as compared to 16 µm for Parker.

To analyze the flow behavior, we plot the volume-weighted flow speed distribution for Parker and Bentheimer samples in Supplementary Figures S3(b) and S3(d), respectively. Both distributions are bimodal, however, with varying peak positions and relative weights. While the first peak occurs at negligible flow speeds, of the order of pm/s, the second peak exhibits flow speeds of the order of µm/s. In Parker, there is a larger fraction of the fluid volume at rest, leading to a small volume-weighted average flow speed of 2 µm/s. In contrast, Bentheimer has a much higher volume-weighted average flow speed of 23 µm/s. The comparison of flow speed distributions reveals that the connection between porosity and permeability is not straightforward: the near-zero-velocity peak represents the volume fraction of the porous medium that, despite being fully connected, does not contribute significantly to permeability.

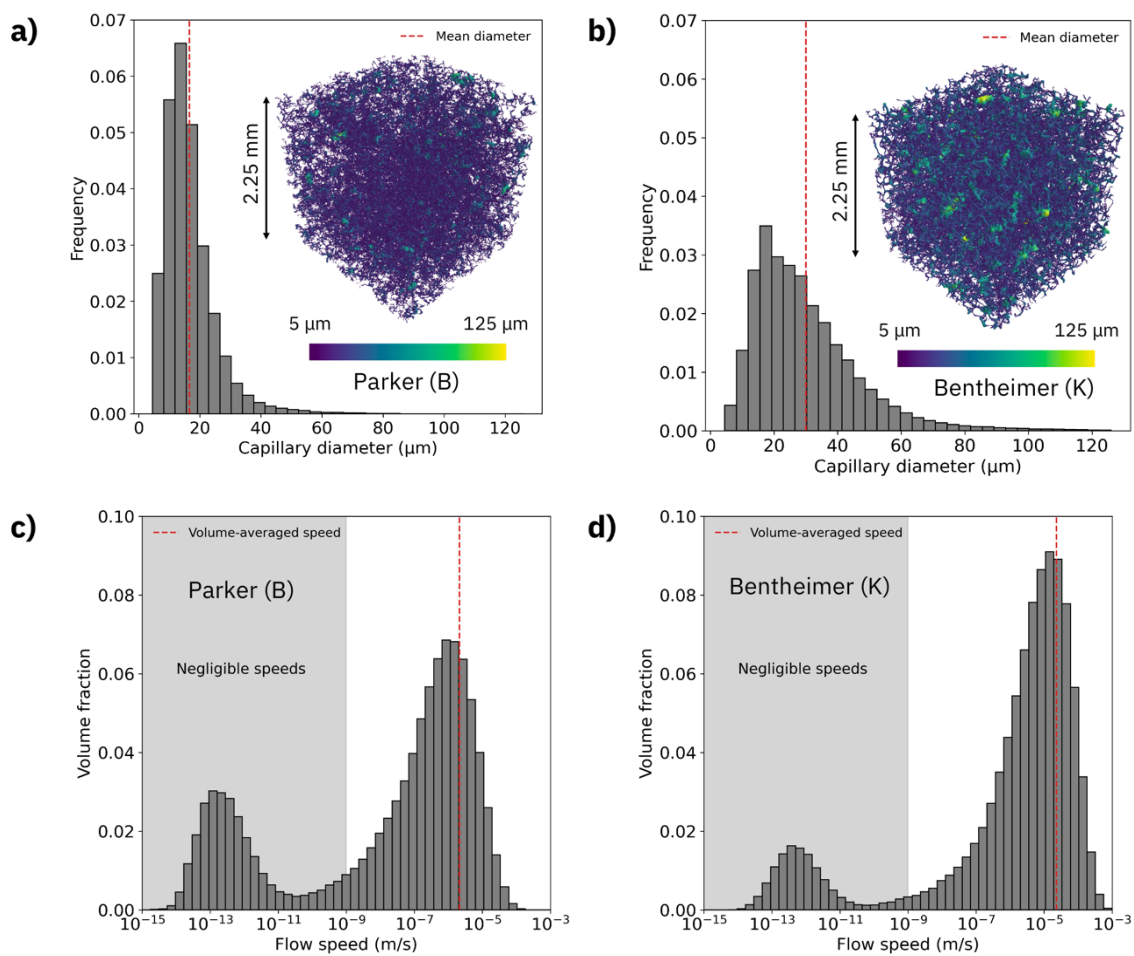

*Supplementary Figure S3: Capillary diameter and flow speed distribution, respectively, obtained by CNM. Capillary diameter distribution of (a) Parker and (b) Bentheimer rock samples investigated in this study, together with visualizations of their capillary network as inset graphs. Darker colors (purple) represent smaller capillary diameters while lighter colors (yellow)*

*represent larger diameters. (c), (d) Volume-weighted flow speed distribution of same the samples shown in (a) and (b), respectively.*

In Supplementary Figure S4, we plot visual maps of the functional dependence of flow speed on capillary diameter obtained for four representative rock samples. All capillary diameter distributions are unimodal, while the flow speed distributions are bimodal. A weak correlation exists between the positions of the peaks, such that the peak at lower flow speed is more prominent for lower diameters and the peak at higher flow speeds is more prominent for higher diameters. Nevertheless, for any given diameter we observe both slower and faster flow speed components in the graphs.

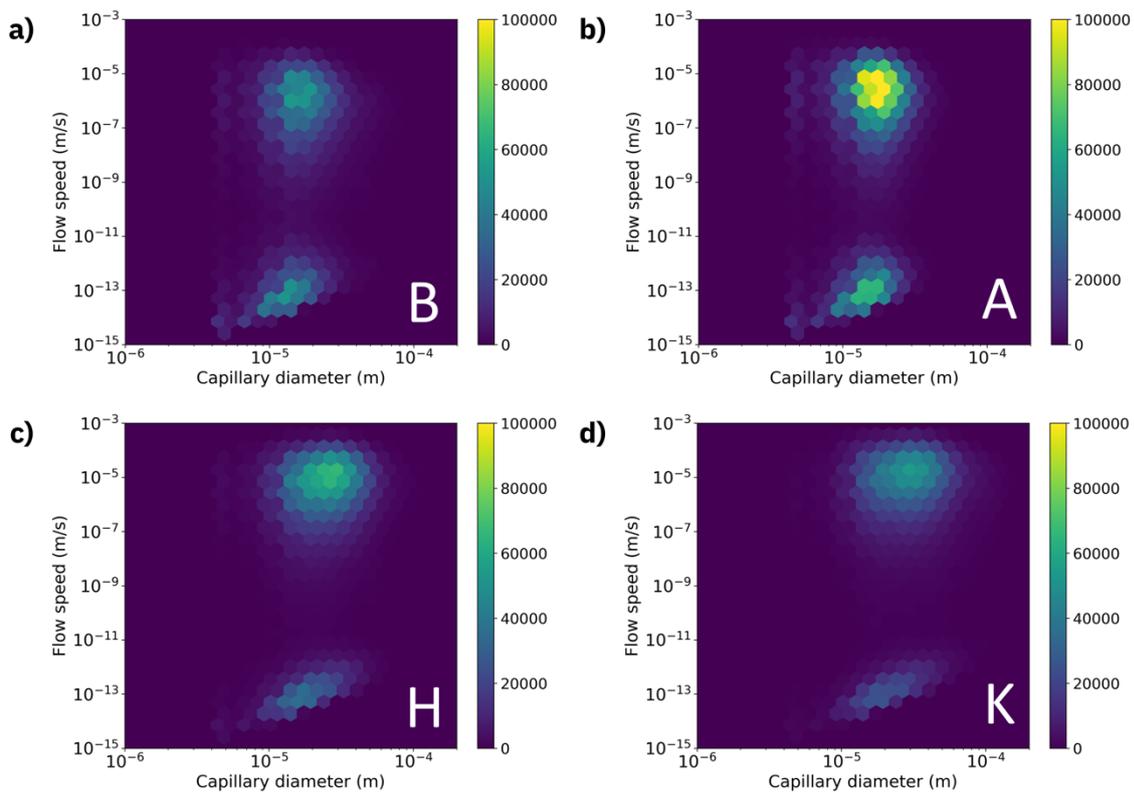

*Supplementary Figure S4: Visualizing flow speeds versus capillary diameters for representative rock samples. 2D-false-color plots depicting (a) the least porous, (b) the least permeable, (c) the most porous and (d) the most permeable sample of this study, respectively.*